\documentclass[12pt]{article}
\usepackage{graphicx}
\usepackage{graphicx}
\newlength{\vshift}
\input epsf.tex
\newlength{\hshift}

\usepackage{amssymb}
\usepackage{amsmath}
\usepackage{amsthm}
\usepackage{amsopn}
\def\uno{\mbox{1 \kern-.59em {\rm l}}}

\def\beq{\begin{equation}}
\def\eeq{\end{equation}}
\def\bea{\begin{eqnarray}}
\def\eea{\end{eqnarray}}

\begin{document}

 \vspace*{3cm}

\begin{center}

{\bf{\Large Nonperturbative Thermodynamic Geometry of Anyon Gas}}

\vskip 4em

{ {\bf Behrouz ~Mirza$^{\dag \S}$} \footnote{
b.mirza@cc.iut.ac.ir}\: and \: {\bf Hosein ~Mohammadzadeh$^{\dag}$
}\footnote{ h.mohammadzadeh@ph.iut.ac.ir}}

\vskip 1em

$^{\dag}$Department of Physics, Isfahan University of Technology,
Isfahan, 84156-83111, Iran

\vskip 1em

$^{\S}$Institute of Theoretical Science, University of Oregon,
Eugene, Oregon 97403-5203, USA
 \end{center}

 \vspace*{1.9cm}

\begin{abstract}
Following our earlier work on the Ruppeiner geometry of an anyon
gas [B. Mirza and H. Mohammadzadeh, Phys. Rev. E {\bf 78,} 021127
(2008)], we will derive nonperturbative thermodynamic curvature of
a two-dimensional ideal anyon gas. At different values of the
thermodynamic parameter space, some unique and interesting
behaviors of the anyon gas are explored. A complete picture of
attractive and repulsive phases of the anyon gas is given.
\end{abstract}

PACS number(s): 05.20.-y, 67.10.Fj
\newpage
\section{Introduction}
The geometrical structure of phase space of statistical
thermodynamics was explicitly studied by Gibbs. The geometrical
thermodynamics was developed by Ruppeiner and Weinhold
{\cite{Ruppeiner1,weinhold1}}. They introduced two sorts of
Riemannian metric structure representing thermodynamic fluctuation
theory, which were related to the second derivative of entropy or
internal energy. This theory represents a new qualitative tool
for the study of fluctuation phenomena. The thermodynamic
curvature has already been calculated for some models whose
thermodynamics are exactly known, where reviews for these models
can be found in \cite{Ruppeiner01,brody2}. Recently, this approach
has been utilized to study the thermodynamics of black holes
\cite{Aman1,Ruppeiner,Mirza1,alvarez}. The thermodynamic
curvature of the ideal classical gas is zero and it could be a
criterion for statistical interaction of the system
\cite{Ruppeiner1,Nulton}. Janyszek and Mruga{\l}a worked out the
thermodynamic curvature for ideal Fermi and Bose gases and
reported that the sign of the thermodynamic curvature is always
different for ideal Fermi and Bose gases. It was argued that the
scalar curvature could be used to show that fermion gases were
more stable than boson gases \cite{Mrugala2}. Also, phase
transition properties of van der Waals gas and some other
thermodynamic models have been considered and it has been shown
that the singular point of the thermodynamic curvature coincides
with the critical point of the system \cite{brody,Janke}.
Recently thermodynamic curvature of the classical limit of the
anyon gas has been worked out \cite{Mirza2}.
 For a two dimensional system, the statistical distribution may interpolate
between fermions and bosons and respects a fractional exclusion
principle \cite{Haldane}. Particles with the new statistics were
named "anyons" by Wilczek \cite{wilczek1}. The thermodynamic
properties of systems with fractional statistical particles or
anyons have been considered and some factorizable properties of
these systems were introduced by Huang \cite{wung1}. It has been
shown that the thermodynamic quantities of a free anyon gas may
be factorized to ideal Bose and Fermi gases \cite{wung2,wung3}.
Using Huang's factorized method, we will explore the
thermodynamics curvature of the anyon gas in the full physical
range.

The outline of this paper is as follows. In Sec. 2, the
thermodynamic properties of anyons are summarized and the internal
energy of the anyon gas is derived. In Sec. 3, the factorizable
properties of fractional statistical particles are collected and
the internal energy and the particle number of the system are
evaluated with respect to the internal energy and the particle
number of  fermion and boson gases. In Sec. 4, the metric of the
parameter space of this system is obtained and, finally, the
thermodynamic curvature of the anyon gas in the full physical
range is evaluated and its properties are investigated.

\section{Thermodynamic properties of the ideal gas of fractional statistical particles }
Particles with fractional statistics or anyons and their
thermodynamic properties have been the subject of research by a
number of authors \cite{Haldane,Leinaas,wilczek,Wu}. Fractional
exchange statistics arises when the many-body wave function of a
system of indistinguishable particles is allowed to acquire an
arbitrary phase $e^{i\pi\alpha}$ upon an adiabatic exchange
process of two particles. Here, $\alpha$ is the so-called
statistical parameter, interpolating between $\alpha =0$ (bosons)
and $\alpha=1$ (fermions). Such an exchange produces a nontrivial
phase only if the configuration space of the collection of
particles under study possesses a multiply connected topological
structure. Therefore, fractional exchange statistics is usually
restricted to two spatial dimensions, $d=2$. However, fractional
exchange statistics can be formalized, to some extent, also in
$d=1$. A different concept of fractional statistics, namely,
fractional exclusion statistics, is based on the structure of the
Hilbert space, rather than the configuration space, of the
particle assembly, and is therefore not restricted to $d\leq 2$
\cite{Haldane,murthy1,murthy2,nayak,pellegrino,prl,Chi}. The
statistical distribution function of anyons has been derived by
Wu using  Haldane's fraction exclusion statistics \cite{Wu},
    \bea
    \label{n}
    n_{i}=\frac{1}{\textit{w}(e^{(\epsilon_{i}-\mu)/kT})+\alpha}
    \eea
where, the function $\textit{w}(\zeta)$ satisfies the functional
equation
    \bea\label{w}
    \textit{w}(\zeta)^{\alpha}[1+\textit{w}(\zeta)]^{1-\alpha}=\zeta\equiv
    e^{(\epsilon-\mu)/kT}
    \eea
and $\alpha$ is the fractional statistical parameter.  The
functional equation for $\textit{w}(\zeta)$ can be solved
analytically only in a few special cases \cite{Aoyama}. Equation
(\ref{w}) yields the correct solutions for two familiar cases:
bosons $(\alpha=0)$, $\textit{w}(\zeta)=\zeta-1$ and fermions
$(\alpha=1)$, $\textit{w}(\zeta)=\zeta$. We can also solve Eq.
(\ref{w}) in the classical limit where
$\exp[(\epsilon-\mu)/kT]\gg1]$,
    \bea
    \label{cl}\textit{w}(\zeta)=\zeta+\alpha-1,
    \eea
    \bea
    n_{i}=\frac{1}{e^{(\epsilon_{i}-\mu)/kT}+2\alpha-1}.
   \eea
Deviation from the classical limit and a more general solution of
Eq. (\ref{w}) is given by the following function
   \bea
    \label{w1}\textit{w}(\zeta)=\zeta+\alpha-1+\frac{c_1}{\zeta}+\frac{c_2}{\zeta^{2}}
    +\frac{c_3}{\zeta^3}+\cdots,
   \eea
where, the constant coefficient $c_{1},c_{2},...$ can be
evaluated on the condition that at each order of $\zeta$,
$\textit{w}(\zeta)$ satisfies Eq. (\ref{w}). The following
solutions are obtained order by order:
   \bea
    c_1&=&\frac{1}{2}\alpha(1-\alpha),\nonumber\\
    c_2&=&\frac{1}{3}\alpha(1-\alpha)(1-2\alpha),\nonumber\\
    c_3&=&\frac{1}{8}\alpha(1-\alpha)(1-3\alpha)(2-3\alpha),\\
    \vdots\nonumber\\
    c_m&=&-\frac{1}{[(m+1)!]m}\prod_{i=0}^{m}(i-m\alpha).\nonumber
    \eea
Now, it is straightforward to obtain the internal energy and
 the particle number of the anyon gas in the classical limit and
perturbatively with a small deviation from the classical limit by
the following relation
 \bea
 \label{UN}U&=&\sum_{i}n_{i}\epsilon_{i},\nonumber\\
 N&=&\sum_{i}n_{i}.
 \eea
In the thermodynamic limit and for two dimensional momentum spaces
of non-relativistic free anyons with a mass $m$, the summation can
be replaced with the following integral:
    \bea \sum_{i}\longrightarrow
    \label{LR}\frac{V}{h^{2}}2\pi m \int_{0}^{\infty} d\epsilon.
    \eea
It should be noted that for obtaining above equations a
 free-particle energy-impulse relation has been used \cite{Wu}.
For a small deviation from the classical limit, we may use the
first correction in Eq. (\ref{w1}) and obtain the internal energy
and the particle number as is presented in \cite{Mirza2}.  We can
use the other correction terms and get far from the classical
limit perturbatively. It is obvious that this procedure does not
yield a
 nonperturbative information. We will review a
nonperturbative  approach based on the factorizable properties of
thermodynamic quantities of the anyon gas in the next section
\cite{wung1,wung2,wung3}.


\section{Factorizable thermodynamic quantities of the anyon gas}
 Huang  showed that a system of free anyons is equivalent to a
system with an $\alpha$ fraction of fermions and a $(1-\alpha)$
fraction of bosons, while the transmutation between the boson and
the  fermion is allowed. The system with boson-fermion
transmutation can be regarded as the ensemble average of $M$
systems, which are classified as  fermions (there are $ \alpha
 M $)  and bosons [there are $(1-\alpha)M]$, and each one of
both cases is equal to $N$-fermion (boson) gas existing in the
volume $V$ and  the pressure $P$. Therefore, the ensemble average
for the thermodynamic quantity $Q_{N}(\alpha)$ of the system with
the boson-fermion transmutation, and thereupon, the anyon system
can be factorized as \cite{wung2}
 \bea
 Q_{N}(\alpha)=\alpha Q_{N}(1)+(1-\alpha)Q_{N}(0),
 \eea
 where, $Q_{N}(\alpha)$ refers to the thermodynamic quantities of the
 anyon system, while $Q_{N}(1)$ and $Q_{N}(0)$ are related to the
 thermodynamic quantities of fermion and boson gases, respectively.
 We will write the above equation in the following simpler form
 \bea
 Q_{a}=\alpha Q_{f}+(1-\alpha)Q_{b}.
 \eea
 Therefore, we can evaluate the internal energy and the particle number of the anyon gas as a
 composition of the internal energy and the particle number of fermion and boson gases,
 while the particle numbers of anyon, fermion, and boson gases are the
 same.
  \bea
  \label{UNA}U_{a}&=&\alpha U_{f}+(1-\alpha)U_{b},\nonumber\\
  N_{a}&=&\alpha N_{f}+(1-\alpha)N_{b}.
  \eea
Wu  also derived the relation \cite{Wu}
 \bea
  \frac{\mu_{a}}{kT}=\alpha\frac{h^{2}}{2\pi V m}\frac{N}{kT}+\ln[1-\exp(-\frac{h^{2}}{2\pi V m}\frac{N}{kT})]\label{fu}.
 \eea
We can rewrite the above equation for the boson and the fermion
cases, with $\alpha=0$ and $\alpha=1$. It should be noted that
 \bea
 \label{nu} N=N_a=N_f=N_b.
 \eea
 It can be easily shown that
 \bea
 \mu_{a}=\alpha\mu_{f}+(1-\alpha)\mu_{b},
 \eea
where, $\mu_a$, $\mu_f$, and $\mu_b$ denote the chemical potential
for anyon, fermion, and boson cases, respectively, which is
consistent with the factorizable property. Also one can derive
 \bea
 \label{z}z_a=z_{f}^{\alpha}z_{b}^{(1-\alpha)},
 \eea
where
 \bea
  \label{fug}z_{a}&=&\exp(\mu_{a}/kT)=e^{\alpha N_{a}\beta/A
  }(1-e^{-N_{a}\beta/A}),\nonumber\\
  z_{f}&=&\exp(\mu_{f}/kT)=e^{N_{f}\beta/A}(1-e^{-N_{f}\beta/A}),\\
  z_{b}&=&\exp(\mu_{b}/kT)=(1-e^{-N_{b}\beta/A})\nonumber
  \eea
  are the fugacity of anyon, fermion, and
boson cases, respectively, and $A=\frac{2\pi V m}{h^{2}}$ and
$\beta=1/kT$.

Albeit it is impossible to solve the functional equation (\ref{w})
for all values of $\alpha$ and in the full physical range, the
factorizable properties of thermodynamic quantities make it
possible to obtain the internal energy and the particle number of
the anyon gas \cite{wung1,wung2,wung3}.  The statistical
distribution function of fermion and boson cases are given by
 \bea
 (n_{i})_{f}=\frac{1}{e^{(\epsilon_{i}-\mu)/kT}+1}\nonumber\\
 (n_{i})_{b}=\frac{1}{e^{(\epsilon_{i}-\mu)/kT}-1}.
 \eea
Subsequently, the internal energy and the particle number of
fermion and boson gases can be evaluated as in the following
 \bea
    \label{UNB}U_{f}&=&\frac{2\pi V m}{h^{2}}\beta^{-2}\int_{0}^{\infty}\frac{\epsilon
    d\epsilon}{e^{(\epsilon-\mu_{f})/kT}+1}=-A\beta^{-2} Li_{2}(-z_{f}),\nonumber\\
    N_{f}&=&\frac{2\pi V m}{h^{2}}\beta^{-1}\int_{0}^{\infty}\frac{
    d\epsilon}{e^{(\epsilon-\mu_{f})/kT}+1}=A\beta^{-1} \ln(1+z_{f})\label{ff},
 \eea
 \bea
 U_{b}&=&\frac{2\pi V m}{h^{2}}\beta^{-2}\int_{0}^{\infty}\frac{\epsilon
    d\epsilon}{e^{(\epsilon-\mu_{b})/kT}-1}=A\beta^{-2} Li_{2}(z_{b}),\nonumber\\
 N_{b}&=&\frac{2\pi V m}{h^{2}}\beta^{-1}\int_{0}^{\infty}\frac{
    d\epsilon}{e^{(\epsilon-\mu_{b})/kT}-1}=-A\beta^{-1} \ln(1-z_{b})\label{bb}.
 \eea
Therefore, the internal energy and the particle number of the
anyon gas will be
 \bea
 U_{a}&=&A\beta^{-2}[-\alpha Li_{2}(-z_{f})+(1-\alpha)Li_{2}(z_{b})],\nonumber\\
 N_{a}&=&A\beta^{-1}[\alpha\ln(1+z_{f})-(1-\alpha)\ln(1-z_{b}]\label{aa},
 \eea
where, $Li_{n}(x)$ denotes the polylogarithm function. The
relations between the thermodynamic quantities of the anyon gas
and the composition of the thermodynamic quantities of fermion and
boson gases lead to some interesting and non-trivial integral
equalities that have been presented in the Appendix.
\section{Thermodynamic curvature of Anyon gas}
Ruppeiner geometry is based on the entropy representation, where
we denote the extended set of $n+1$ extensive variables of the
system by $X=(U,N^{1},...,V,...,N^{r})$, while
 Weinhold worked in the energy representation in which the
extended set of $n+1$ extensive variables of system are denoted
by $Y=(S,N^{1},...,V,...,N^{r})$ \cite{Ruppeiner01}. It should be
noted that we can work in any thermodynamic potential
representation that is the Legendre transform of the entropy or
the internal energy. The metric of this representation may be the
second derivative of the thermodynamic potential with respect
 to intensive variables; for example, the thermodynamic potential
$\Phi$ which is defined as,
    \bea
    \Phi=\Phi(\{F^{i}\}),
    \eea
where, $F=(1/T,-\mu^{1}/T,...,P/T,...,-\mu^{r}/T)$. $\Phi$ is the
Legendre transform of entropy with respect to the extensive
parameter $X^{i}$,
    \bea
    F^{i}=\frac{\partial S}{\partial X^{i}}.
    \eea
The metric in this representation is given by
    \bea
    g_{ij}=\frac{\partial^{2}\Phi}{\partial F^{i}\partial F^{j}}.
    \eea
Janyszek and Mruga{\l}a used the partition function to introduce
the metric geometry of the parameter space \cite{Mrugala2},
    \bea\label{M1} g_{ij}=\frac{\partial^{2}\ln
    Z}{\partial \beta^{i}\partial\beta^{j}}
    \eea
where, $\beta^{i}=F^{i}/k$ and $Z$ is the partition function.

According to Eqs. (\ref{ff}),(\ref{bb}), and (\ref{aa}), the
parameter space of  ideal fermion, boson and anyon gases are
$(\beta,\gamma_{f})$, $(\beta,\gamma_{b})$, and
$(\beta,\gamma_{a})$, respectively, where $ {\beta} = {1 / kT}$
and $\gamma_i = -\mu_i/ kT $ . For computing the thermodynamic
metric, we select one of the extended variables as the constant
system scale. We will implicitly pick $V$ in working with the
grand canonical ensemble \cite{Mrugala2}. We can evaluate the
metric elements of fermion $(F_{ij})$, boson $(B_{ij})$, and
anyon gases $(A_{ij})$ by the definition of metric in Eq.
({\ref{M1}). The metric elements of the thermodynamic space of an
ideal fermion gas are given by
    \bea
    F_{\beta\beta}&=&\frac{\partial^{2}\ln Z_{f}}{\partial
    \beta^{2}}=-(\frac{\partial U_{f}}{\partial\beta})_{\gamma_{f}}=-2\beta^{-3}Li_{2}(-z_{f}),\nonumber\\
    F_{\beta\gamma_{f}}&=&F_{\gamma_{f}\beta}=\frac{\partial^{2}\ln Z_{f}}{\partial
    \beta\partial\gamma_{f}}=-(\frac{\partial U_{f}}{\partial
    \gamma_{f}})_{\beta}=\beta^{-2}\ln(1+z_{f}),\\
    \label{G1} F_{\gamma_{f}\gamma_{f}}&=&\frac{\partial^{2}\ln Z_{f}}{\partial
    \gamma_{f}^{2}}=-(\frac{\partial
    N_{f}}{\partial\gamma_{f}})_{\beta}=\beta^{-1}\frac{z_{f}}{1+z_{f}}.\nonumber
    \eea
In the same manner, the metric elements of the thermodynamic space
of an ideal boson gas are given by
    \bea
    B_{\beta\beta}&=&\frac{\partial^{2}\ln Z_{b}}{\partial
    \beta^{2}}=-(\frac{\partial U_{b}}{\partial\beta})_{\gamma_{b}}=2\beta^{-3}Li_{2}(z_{b}),\nonumber\\
    B_{\beta\gamma_{b}}&=&B_{\gamma_{b}\beta}=\frac{\partial^{2}\ln Z_{b}}{\partial
    \beta\partial\gamma_{b}}=-(\frac{\partial U_{b}}{\partial
    \gamma_{b}})_{\beta}=-\beta^{-2}\ln(1-z_{b}),\\
    \label{G2} B_{\gamma_{b}\gamma_{b}}&=&\frac{\partial^{2}\ln Z_{b}}{\partial
    \gamma_{b}^{2}}=-(\frac{\partial
    N_{b}}{\partial\gamma_{b}})_{\beta}=\beta^{-1}\frac{z_{b}}{1-z_{b}}.\nonumber
    \eea
For simplicity, we have set the constant $A=1$. By using the
factorizable properties, the metric elements of thermodynamic
space of an ideal anyon gas can be derived as follows:
    \bea
     \label{vv} A_{\beta\beta}&=&\frac{\partial^{2}\ln Z_{a}}{\partial
    \beta^{2}}=-(\frac{\partial U_{a}}{\partial\beta})_{\gamma_{a}}=-\frac{\partial}{\partial\beta}(\alpha U_{f}+(1-\alpha) U_{b})\nonumber\\
    &=&\alpha F_{\beta\beta}+(1-\alpha) B_{\beta\beta}=2\beta^{-3}(-\alpha Li_{2}(-z_{f})+(1-\alpha)Li_{2}(z_{b})),\nonumber\\
    A_{\beta\gamma_{a}}&=&\frac{\partial^{2}\ln Z_{a}}{\partial
    \gamma_{a}\partial\beta}=-(\frac{\partial N_{a}}{\partial\beta})_{\gamma_{a}}=-\frac{\partial}{\partial\beta}[\alpha N_{f}+(1-\alpha) N_{b}]\nonumber\\
    &=&\alpha F_{\beta\gamma_{f}}+(1-\alpha) B_{\beta\gamma_{b}}=\beta^{-2}[\alpha\ln(1+z_{f})-(1-\alpha)\ln(1-z_{b})],\\
     A_{\gamma_{a}\gamma_{a}}&=&\frac{\partial^{2}\ln Z_{a}}{\partial
    \gamma_{a}^{2}}=-(\frac{\partial N_{a}}{\partial\gamma_{a}})_{\beta}=-1/(\frac{\partial\gamma_{a}}{\partial N_{a}})_{\beta}\nonumber\\
    &=&\frac{1}{\alpha/F_{\gamma_{f}\gamma_{f}}+(1-\alpha)/B_{\gamma_{b}\gamma_{b}}}
    =\beta^{-1}\frac{-z_{f}z_{b}}{2\alpha z_{f}z_{b}+\alpha z_{b}-\alpha z_{f}-z_{f}z_{b}+z_{f}},\nonumber
    \eea
To obtain the last equation we use Eq. (\ref{z}) and differentiate
with respect to $N$; the particle number of system,
  \bea
  (\frac{\partial\mu_{a}}{\partial N_{a}})_{\beta}=\alpha(\frac{\partial\mu_{f}}{\partial N_{f}})_{\beta}+(1-\alpha)(\frac{\partial\mu_{b}}{\partial
  N_{b}})_{\beta}.\label{op}
  \eea
We consider a system with two thermodynamic degrees of freedom
and,
 therefore, the dimension of the thermodynamic surface or the parameter
space is equal to $2$ ($D=2$). Thus, the scalar curvature is given
by
    \bea
    R=\frac{2}{\det g} R_{1212.}
    \eea
 Janyszek and Mruga{\l}a demonstrated
\cite{Mrugala3} that if the metric elements  are written purely as
the second derivatives of a certain thermodynamic potential, the
thermodynamic curvature may then be written in terms of the second
and the third derivatives. The sign convention for $R$ is
arbitrary, so $R$ may be either negative or positive for any
case. Our selected sign convention is the same as that of Janyszek
and Mruga{\l}a, but opposite from \cite{Ruppeiner01}. In
 two-dimensional spaces, the Ricci scalar is defined by

 \bea
R=\frac{2\left|
         \begin{array}{ccc}
           g_{\beta\beta} & g_{\gamma\gamma} & g_{\beta\gamma} \\
           g_{\beta\beta,\beta} & g_{\gamma\gamma,\beta} & g_{\beta\gamma,\beta} \\
           g_{\beta\beta,\gamma} & g_{\gamma\gamma,\gamma} & g_{\beta\gamma,\gamma} \\
         \end{array}
       \right|}{{\left|
                  \begin{array}{cc}
                    g_{\beta\beta} & g_{\beta\gamma} \\
                    g_{\beta\gamma} & g_{\gamma\gamma} \\
                  \end{array}
                \right|}^{2}
       }.
    \eea
Using the following equations for a fermion gas:
    \bea
    F_{\beta\beta,\beta}&=&6\beta^{-4}Li_{2}(-z_{f}),\nonumber\\
    F_{\beta\beta,\gamma_{f}}&=&F_{\beta\gamma_{f},\beta}=-2\beta^{-3}\ln(1+z_{f}),\nonumber\\
    F_{\gamma_{f}\gamma_{f},\beta}&=&F_{\beta\gamma_{f},\gamma_{f}}=-\beta^{-2}\frac{z_{f}}{1+z_{f}},\nonumber\\
    F_{\gamma_{f}\gamma_{f},\gamma_{f}}&=&-\beta^{-1}\frac{z_{f}}{(1+z_{f})^{2}},
    \eea
and the following equations for the boson gas
    \bea
    B_{\beta\beta,\beta}&=&-6\beta^{-4}Li_{2}(z_{b}),\nonumber\\
    B_{\beta\beta,\gamma_{b}}&=&B_{\beta\gamma_{b},\beta}=2\beta^{-3}\ln(1-z_{b}),\nonumber\\
    B_{\gamma_{b}\gamma_{b},\beta}&=&B_{\beta\gamma_{b},\gamma_{b}}=-\beta^{-2}\frac{z_{b}}{1-z_{b}},\nonumber\\
    B_{\gamma_{b}\gamma_{b},\gamma_{b}}&=&-\beta^{-1}\frac{z_{b}}{(1-z_{b})^{2}},
    \eea
we can obtain the following equations for an anyon gas
    \bea
    A_{\beta\beta,\beta}&=&\alpha
    F_{\beta\beta,\beta}+(1-\alpha)B_{\alpha\alpha,\alpha}\nonumber\\
    &=&6\beta^{-4}[\alpha Li_{2}(-z_{f})-(1-\alpha)Li_{2}(z_{b})],\nonumber\\
    A_{\beta\beta,\gamma_{a}}&=&G_{\beta\gamma_{a},\beta}=\alpha
    F_{\beta\beta,\gamma_{f}}+(1-\alpha)B_{\beta\beta,\gamma_{b}}\nonumber\\
    &=&2\beta^{-3}[-\alpha\ln(1+z_{f})+(1-\alpha)\ln(1-z_{b})],\nonumber\\
    A_{\gamma_{a}\gamma_{a},\beta}&=&G_{\beta\gamma_{a},\gamma_{a}}=
    A_{\gamma_{a}\gamma_{a}}^{2}\{\alpha\frac{F_{\beta\gamma_{f},\gamma_{f}}}{F_{\gamma_{f}\gamma_{f}}^{2}}
    +(1-\alpha)\frac{B_{\beta\gamma_{b},\gamma_{b}}}{B_{\gamma_{b}\gamma_{b}}^{2}}\}\nonumber\\
    &=&-\beta^{-2}\frac{z_{f}z_{b}}{2\alpha z_{f}z_{b}
    +\alpha z_{b}-\alpha z_{f}-z_{f}z_{b}+z_{f}},\nonumber\\
    A_{\gamma_{a}\gamma_{a},\gamma_{a}}&=&A_{\gamma_{a}\gamma_{a}}^{3}\{\alpha\frac{F_{\gamma_{f}\gamma_{f},\gamma_{f}}}{F_{\gamma_{f}\gamma_{f}}^{3}}
    +(1-\alpha)\frac{B_{\gamma_{b}\gamma_{b},\gamma_{b}}}{B_{\gamma_{b}\gamma_{b}}^{3}}\}\nonumber\\
    &=&-\beta^{-1}\frac{z_{f}z_{b}[\alpha z_{b}^{2}(1+z_{f})+\alpha z_{f}^{2}(-1+z_{b})+z_{f}^{2}(1-z_{b})]}{(2\alpha z_{f}z_{b}
    +\alpha z_{b}-\alpha z_{f}-z_{f}z_{b}+z_{f})^{3}}.
    \eea
The third equations is obtained from Eq. (\ref{vv}),
 \bea
 A_{\gamma_{a}\gamma_{a},\beta}=(\frac{\partial
 A_{\gamma_{a}\gamma_{a}}}{\partial\beta})_{\gamma_{a}}=\frac{\partial}{\partial\beta}[\alpha/F_{\gamma_{f}\gamma_{f}}
 +(1-\alpha)/B_{\gamma_{b}\gamma_{b}}]^{-1}
 \eea
and the last equation comes from  differentiating Eq. (\ref{op})
with respect to $N$ and also by using the  following equation:
 \bea
 \frac{\partial^{2}\gamma_a}{\partial N_{a}^{2}}=-\frac{1}{(\frac{\partial
 N_a}{\partial\gamma_a})^{3}}\frac{\partial^{2}N_{a}}{\partial\gamma_{a}^{2}}.
 \eea
  Now, we can calculate
the thermodynamic curvature for the ideal fermion, boson, and
anyon gases. The Ricci scalar for fermion and boson gases are
given by,
 \bea
 R_{f}=-\frac{4\beta
 z_{b}(Li_{2}(-z_{f})\ln(1+z_{f})-2z_{f}Li_{2}(-z_{f})-\ln^{2}(1+z_{f})-z_{f}\ln^{2}(1+z_{f}))}
 {(2z_{f}Li_{2}(-z_{f})+\ln^{2}(1+z_{f})+z_{f}\ln^{2}(1+z_{f}))^{2}},
 \eea
 \bea
 R_{b}=-\frac{4\beta
 z_{b}(Li_{2}(z_{b})\ln(1-z_{b})+2z_{b}Li_{2}(z_{b})-\ln^{2}(1-z_{b})+z_{b}\ln^{2}(1-z_{b}))}
 {(2z_{b}Li_{2}(z_{b})-\ln^{2}(1-z_{b})+z_{b}\ln^{2}(1-z_{b}))^{2}}.
 \eea
   \begin{figure}
    \center
    \includegraphics[width=10cm]{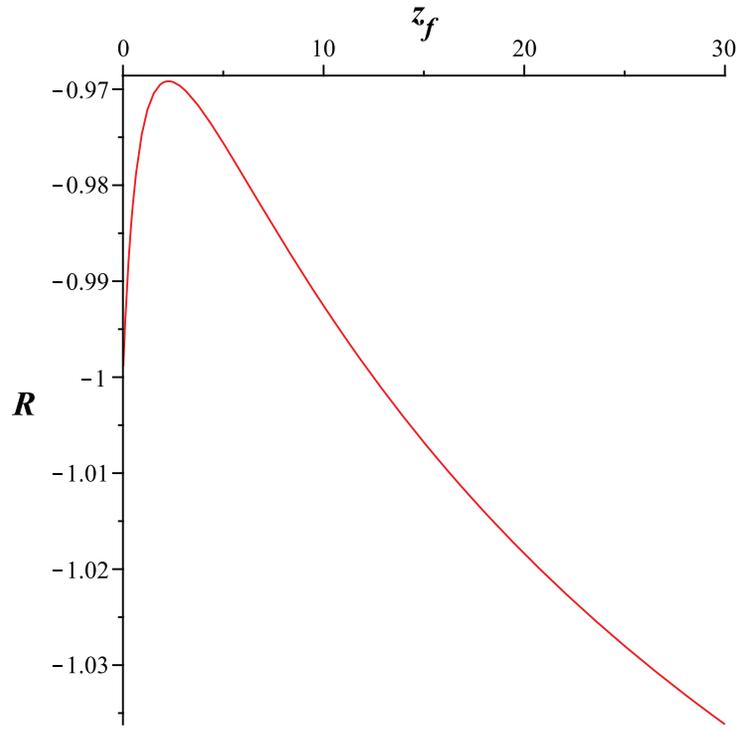}\\
    \caption{(Color online) The thermodynamic curvature of a fermion gas as a function of $z_{f}$
    for an isotherm ($\beta=1$). }\label{rf}
   \end{figure}

   \begin{figure}
    \center\includegraphics[width=8cm]{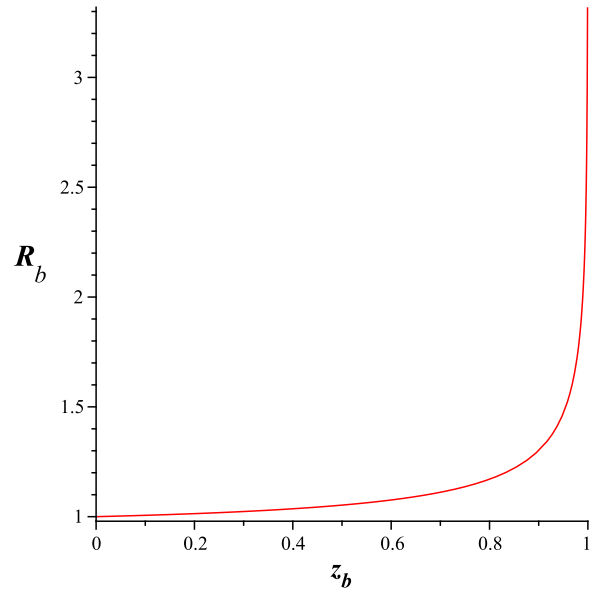}\\
    \caption{(Color online) The thermodynamic curvature of a boson gas as a function of $z_{b}$ for an isotherm ($\beta=1$).}\label{rb}
   \end{figure}
 The thermodynamic curvature of the ideal fermion and boson gases are depicted in Figs.  \ref{rf} and \ref{rb}. The Ricci scalars for  fermion and
 boson gases are negative and positive, respectively. The thermodynamic curvature of a boson gas also has a
 singularity at $z_{b}=1$. The Bose-Einstein condensation phase
 transition occurs at this point in the three-dimensional space
 \cite{Mrugala2}. In the  two-dimensional space, there is no temperature
 below which the ground state can be said to be macroscopically
 occupied in comparison to the excited states. Therefore, as is
 well known, no Bose-Einstein condensation occurs in the two-dimensional space \cite{May,Lee}. But the singular property of the
 thermodynamic curvature at $z_{b}=1$ in the  two-dimensional space has remained
 from the higher dimension. One can realize from Figure
 \ref{rf} that the thermodynamic curvature of a fermion gas has a
maximum point. As shown in \cite{Mrugala2,janyszek}, we may
consider the thermodynamic curvature as a measure of the
stability of the system: the bigger the value of $R$, the less
stable is the system. This interpretation of stability measures
the looseness of the system to fluctuations and does not refer to
the fact that the metric is definitely positive. Therefore, the
maximum point of thermodynamic curvature of the fermion gas
coincides with the less stable situation. The thermodynamic
curvature of the anyon gas is intricate and we will explore it in
the following sections.
\subsection{Fixed temperature}
In the following sections, we are going to investigate the
thermodynamic curvature of an anyon gas for an isotherm; hence, we
set $\beta=1$. Therefore, the thermodynamic curvature will be a
function of $\alpha$ and $z_{a}$ ($z_a$ is a function of $z_{f}$
and $z_{b}$).

\subsubsection{Thermodynamic curvature as a function of $\alpha$ and dual points}
We select values of anyon fugacity in the classical limit and get
far from that limit. The result is depicted in Fig. \ref{f1}
which shows the thermodynamic curvature as a function of $\alpha$
for different values of fugacity.  It is obvious that in the
classical limit (small values of fugacity) the thermodynamic
curvature has two different signs. It is positive for
$\alpha<\frac{1}{2}$ while it is negative for
$\alpha>\frac{1}{2}$. The sign of thermodynamic curvature changes
at $\alpha=\frac{1}{2}$ and the anyon gas behaves like an ideal
classical gas. Deviation from the classical limit moves the zero
point of the thermodynamic curvature from $\alpha=\frac{1}{2}$ to
the lower values \cite{Mirza2}. Unique and interesting phenomena
appear at $z_{a}\geq 1$. The thermodynamic curvature for
$z_{a}=1$ goes to infinity at $\alpha=0$ (boson gas), where in the
higher dimensions the Bose-Einstein condensation occurs. For
$z_{a}>1$, the thermodynamic curvature has a maximum point. From
Eqs. (\ref{ff}) and (\ref{bb}), one can find that the particle
number of boson (fermion) gas for an isotherm is convex down (up)
functions with respect to the fugacity, whereas these functions
for the anyon gas for an isotherm with respect to $z_{a}$ face
with a mutation in curvature of the function and has an
inflection point for some values of $\alpha$. This point for
$\alpha=0$ occurs at $z_{a}=1$ while for $\alpha>0$ it coincides
with the maximum value of
 $R$ with a specified value of $z_{a}>1$. This means that the convexity of these functions for any
fixed $\alpha$ changes for a special value of $z_{a}$ that
coincides to the maximum point of the thermodynamic curvature.
According to the stability interpretation of the value of
thermodynamic curvature, these maximum points may be related to a
 less stable state of the system. It is also interesting to note that we  can find two
different values of $\alpha$ with the same value for the
thermodynamic curvature. At some values of $z_{a}$, we obtain two
values for $\alpha$ with zero curvature which indicates a duality
relation between these points.

   \begin{figure}
    \center\includegraphics[width=10cm]{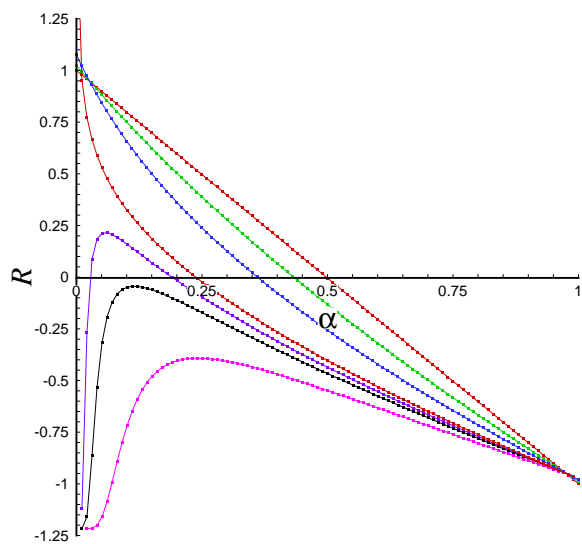}\\
    \caption{(Color online)  The thermodynamic curvature as a function of $\alpha$
    for an isotherm. The values of anyon fugacity have been taken are
    $z_{a}=$ 0.01 [red (upper) line], 0.3 [green (light gray) line], 0.6 (blue line), 1
    [
    red (middle)
    line], 1.1 (blue line), 1.2 [black line] and 1.5 [purple (lower) line].}\label{f1}
   \end{figure}
\subsubsection{Thermodynamic curvature as a function of $z_{a}$}

For $\alpha=\frac{1}{2}$, the full physical range of thermodynamic
curvature has already been considered in \cite{Mirza2}. In this
part, we are going to evaluate the thermodynamic curvature for
fixed values of $\alpha$ and arbitrary values of anyon fugacity.
Figure \ref{f2} represents the thermodynamic curvature of the
anyon gas for an isotherm for three different values of $\alpha$
as a function of anyon fugacity. This figure suggest that for all
values of $\alpha$ (except $\alpha=0$), the thermodynamic
curvature for large values of fugacity may go to fixed negative
values.
\begin{figure}
  \center\includegraphics[width=10cm]{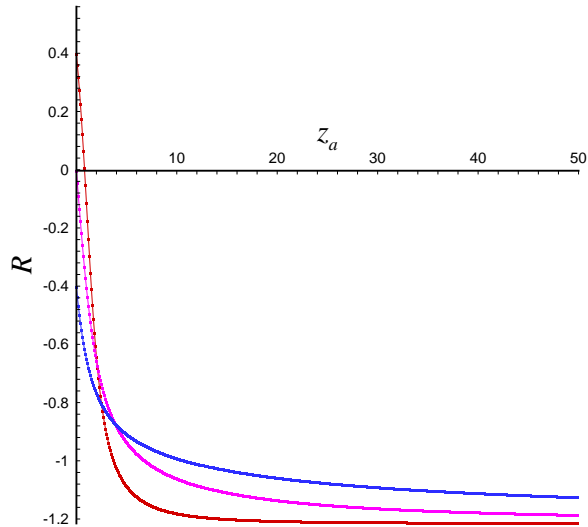}\\
  \caption{ (Color online) The thermodynamic curvature as a function of $z_{a}$ (anyon fugacity) for
  $\alpha=$ 0.3 [red (lower) line], 0.5 [purple (middle) line], 0.7 [blue (upper) line]  and an isotherm in the full physical range.}\label{f2}
\end{figure}
\subsection{Fixed fugacity}
It is straightforward to obtain the thermodynamic curvature at a
fixed fugacity and as a function of $\alpha$. We restrict
ourselves to the more interesting region $z_{a}>1$ and set
$z_{a}=1.5$. The thermodynamic curvature as a function of
$\alpha$ and for three different values of $\beta$ is depicted in
Fig. (\ref{f3}). It is shown that by increasing the value of
$\beta$ or at lower temperatures, the maximum value of the
thermodynamic curvature increases while it also goes toward the
lower values of $\alpha$. This means that at the limit of $T=0$,
the thermodynamic curvature for $\alpha=0$ goes to large values.
Although there is no phase transition in the two dimensional space
we see a  behavior similar to Bose-Einstein condensation.
Actually, the maximum value and sharp changes in the thermodynamic
curvature in the $z_{a}>1$ region, can be interpreted as the
remaining of a phase transition in a higher three-dimensional
world, which is of course the familiar Bose-Einstein condensation.
The dual points with $R=0$ can clearly be identified in fig.
\ref{f3}.
 \begin{figure}
  \center\includegraphics[width=10cm]{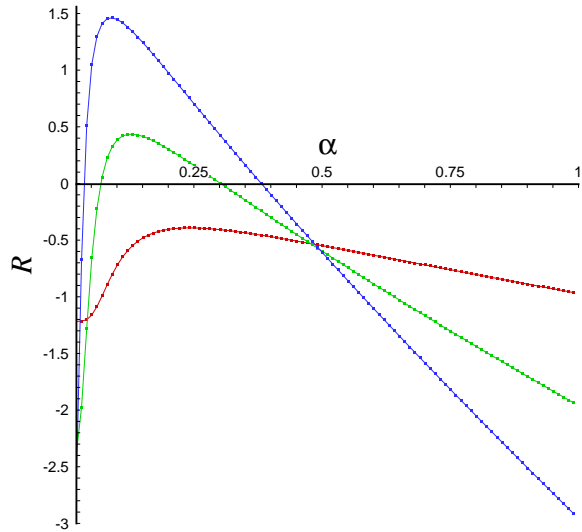}\\
  \caption{(Color online) The thermodynamic curvature as a function of $\alpha$ for
  $\beta=$ 1 [red (lower) line], 2 [green (light gray) line], and 3 [blue (upper maximum) line]  and a fixed fugacity  at $z_{a}=1.5$.}\label{f3}
\end{figure}
\subsection{Fixed particle number}
We can drive the thermodynamic curvature as a function of $\beta$
and $N_{a}$ by substituting the fugacity from (\ref{z}). It has
been shown that at $T=0$, particles of general exclusion
statistics exhibit a Fermi surface \cite{nayak}. This fact
dictates the low temperature thermodynamics of these particles
when the particle number is conserved. Figure (\ref{f4}) shows
the thermodynamic curvature of the anyon gas at different values
of $\alpha$. The particle number has been fixed for simplicity at
$N_{a}=1$. The upper curve coincides with the thermodynamic
curvature of the boson gas ($\alpha=0$) and the lower curve
coincides with the fermion gas ($\alpha=1$). The other curves
show the thermodynamic curvature of the intermediate values of
$\alpha$. It is clear that by increasing the value of $\beta$ or
by  going toward low temperatures,  the thermodynamic curvature
approaches that of a fermion gas, which is consistent with the
result in \cite{nayak}.
\begin{figure}
  \center\includegraphics[width=8cm]{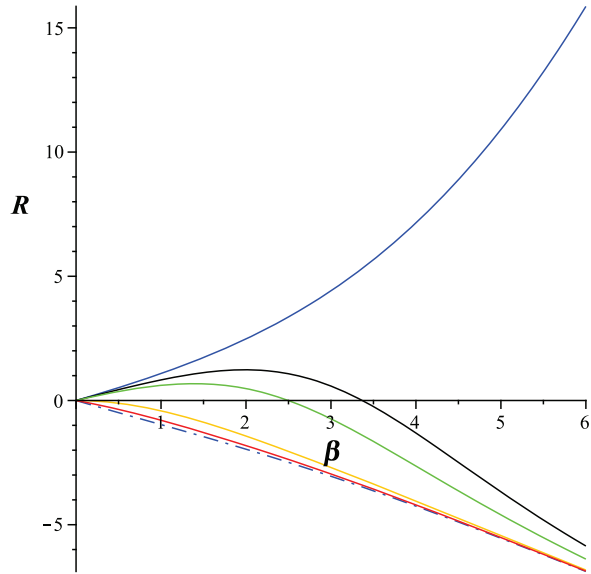}\\
  \caption{(Color online) The thermodynamic curvature as a function of $\beta$ while the particle number is conserved
  ($N_{a}=1$). The solid  and the dashed-dotted blue lines correspond to the boson ($\alpha=0$)
  and the
  fermion ($\alpha=1$) gases respectively. The other lines correspond to
   the intermediate values of fraction parameter $\alpha=$ 0.05 (black line), 0.1 (green line)
   , 0.5 (orange line), and 0.8 (red line).}\label{f4}
\end{figure}

\section{Conclusion}

We derived the non-perturbative thermodynamic curvature of an
ideal anyon gas. It is interesting that, for $z_{a}>1$, there is a
maximum point for the thermodynamic curvature. At low temperatures
and at
 a fixed particle number, the thermodynamic curvature approaches
 that of  a fermion gas, which indicates that, at $T=0$, particles of general
exclusion statistics exhibit a Fermi surface. It is clear from
Fig. \ref{rf} that the thermodynamic curvature of a fermion gas
is negative while it is positive for a boson gas, likewise, the
statistical interaction for a fermion gas is repulsive but it is
attractive for a boson gas. We may propose a unique interpretation
for the thermodynamic curvature of the anyon gas according to its
sign. It has already been shown that in the classical limit,
statistical interaction of an anyon gas can be attractive or
repulsive \cite{Wu,Wung4}. The attractive case corresponds to a
positive curvature and the repulsive one corresponds to a negative
curvature \cite{Mirza2}.  We may identify the attractive and the
repulsive parts for an ideal anyon gas from Figs. 3 and 5. This
work suggests that there may be dual points where we get
equivalent anyon gases but with different $\alpha$'s.

\ \  \ \ \

{\noindent \bf Acknowledgments}

\noindent  B. M. would likes to thank Institute of Theoretical
Science at university of Oregon for hospitality.

\appendix
\section{}
The factorizable property of anyon thermodynamic quantities
enable us to write the internal energy and the particle number of
an anyon gas as a composition of the internal energy and the
particle number of fermion and boson gases while the condition
$N_a=N_f=N_b$ is preserved. Whereas the distribution function of
the anyon gas can be solved analytically for some values of the
fraction parameter $\alpha$, the thermodynamic quantities can be
derived for such values of the fraction parameter. For semions
with $\alpha=1/2$, the statistical distribution function is given
by \cite{Wu},
 \bea
 n_{i}=\frac{1}{\sqrt{1/4+\exp[2(\epsilon_{i}-\mu_{a})/kT]}}=\frac{2}{\sqrt{1+\frac{4}{z_{a}^{2}}\exp(2\epsilon_{i}/kT)}}
 \eea
So, from Eqs. (\ref{UN}) and (\ref{LR}), we get,
 \bea
 U_{a}=A\beta^{-2}\int_{0}^{\infty}\frac{x
 d x}{\sqrt{1/4+\frac{1}{z_{a}^{2}}\exp(2x)}},\nonumber\\
 N_{a}=A\beta^{-1}\int_{0}^{\infty}\frac{
 d x}{\sqrt{1/4+\frac{1}{z_{a}^{2}}\exp(2x)}},
 \eea
and the following equivalent relation can be obtained by using
Eqs.
 (\ref{UNA}) and (\ref{UNB}):
 \bea
 \label{ap}U_{a}=A\beta^{-2}(\frac{1}{2}\int_{0}^{\infty}\frac{x
 d x}{\frac{1}{z_{f}}\exp(x)+1}+\frac{1}{2}\int_{0}^{\infty}\frac{x
 d x}{\frac{1}{z_{b}}\exp(x)-1}),\nonumber\\
 N_{a}=A\beta^{-2}(\frac{1}{2}\int_{0}^{\infty}\frac{
 d x}{\frac{1}{z_{f}}\exp(x)+1}+\frac{1}{2}\int_{0}^{\infty}\frac{
 d x}{\frac{1}{z_{b}}\exp(x)-1}).
 \eea
Therefore, we get the following non-trivial relations,
 \bea
 \label{int} \int_{0}^{\infty}\frac{x
 d x}{\sqrt{1/4+\frac{1}{z_{a}^{2}}\exp(2x)}}=\frac{1}{2}\left(\int_{0}^{\infty}\frac{x
 d x}{\frac{1}{z_{f}}\exp(x)+1}+\int_{0}^{\infty}\frac{x
 d x}{\frac{1}{z_{b}}\exp(x)-1}\right),\nonumber\\
 \int_{0}^{\infty}\frac{
 d x}{\sqrt{1/4+\frac{1}{z_{a}^{2}}\exp(2x)}}=\frac{1}{2}\left(\int_{0}^{\infty}\frac{
 d x}{\frac{1}{z_{f}}\exp(x)+1}+\int_{0}^{\infty}\frac{
 d x}{\frac{1}{z_{b}}\exp(x)-1}\right),
 \eea
where, the value of fugacity for anyon, fermion, and boson gases
must be evaluated from (\ref{fug}) with the condition
$N_a=N_f=N_b$. For example, if we set $z_a=2,  z_f=4.828427125$,
and $z_b=0.8284271247$, the above condition is satisfied and the
integral equalities (\ref{int}) are valid, which can be checked by
MAPLE or MATHEMATICA. The following nontrivial equality can also
be obtained by using Eq. (\ref{vv}), with those values of fugacity
for anyon, fermion, and boson gases, which will satisfy Eq.
(\ref{nu}),
 \bea
 \int_{0}^{\infty}\frac{8z_{a}\exp(2x)d x}{\left[z_{a}^{2}+4\exp(2x)\right]^{3/2}}=
 2\left(\frac{1}{\int_{0}^{\infty}\frac{z_{f}\exp(x)d x}{[\exp(x)+z_{f}]^{2}}}
 +\frac{1}{\int_{0}^{\infty}\frac{z_{b}\exp(x)d x}{[\exp(x)-z_{b}]^{2}}}\right)^{-1}.
 \eea

\end{document}